# Extending the Propagation Distance of a Silver Nanowire Plasmonic Waveguide with a Dielectric Multilayer Substrate

Douguo Zhang[1*], Yifeng Xiang[1], Junxue Chen[2*], Junjie Cheng[3], Liangfu Zhu[1], Ruxue Wang[1], Gang Zou[3], Pei Wang[1], Hai Ming[1], Mary Rosenfeld[4], Ramachandram Badugu[4], and Joseph R. Lakowicz[4]

[1]Institute of Photonics, Department of Optics and Optical Engineering, University of Science and Technology of China, Hefei, Anhui, 230026, P.R. China

[2]School of Science, Southwest University of Science and Technology, Mianyang, Sichuan 621010, P.R. China

[3]CAS Key Laboratory of Soft Matter Chemistry, Department of Polymer Science and Engineering, iChEM, University of Science and Technology of China, Hefei, Anhui 230026, P.R. China

[4]Center for Fluorescence Spectroscopy, Department of Biochemistry and Molecular Biology, University of Maryland School of Medicine, Baltimore, MD 21201, United States

Correspondence and requests for materials should be addressed to D.G. Zhang (Email: dgzhang@ustc.edu.cn) or J. X. Chen (Email: cjxueoptics@163.com).

abstract
**ABSTRACT:**

Chemical synthesized silver nanowires have been proved to be the efficient architecture for Plasmonic waveguides, but the high propagation loss prevents their widely applications. Here, we demonstrate that the propagation distance of the plasmons along the Ag NW can be extended if the Ag NW was placed on a dielectric multilayer substrate containing a photonic band gap, but not placed on a commonly used glass substrate. The propagation distance at 630 nm wavelength can reach 16 μm even that the Ag NW is as thin as 90 nm in diameter. Experimental and simulation results further show that the polarization of this propagating plasmon mode was nearly parallel to the surface of the dielectric multilayer, so it was excited by a transverse-electric polarized Bloch surface wave propagating along a polymer nanowire with diameter at only about 170 nm on the same dielectric multilayer. Numerical simulations were also carried out and consistent with the experiment results. Our work provides a platform to extend the propagation distance of plasmonic waveguide and also for the integration between photonic and plasmonic waveguides on the nanometre scale.

Silver nanowires (Ag NWs) synthesized using wet chemistry approaches have unique properties, such as their single crystallinity and atomic surface smoothness, which makes them particularly useful for optical confinement applications below the diffraction limit and for the guiding of light on the nanometre scale. [1, 2] Each Ag NW functions as a plasmonic waveguide, meaning that surface plasmons (SPs) can be excited and propagate along the NW. [3–5] Till now, chemically synthesized crystalline Ag NWs show low scattering losses for wave guiding and are widely used for the development of nanometre-scale devices. [6] For example, Ag NWs can be assembled to form branched structures to carry out the functions of routers and logic gates, [7–10] to form hybrid photon–plasmon nanowire lasers, [11] and to form hybrid nanophotonic components and circuits with subwavelength confinement and reduced losses. [12]

Ag NWs are very thin and soft and have a large aspect ratio; hence they are placed on solid substrates for practical applications. Typically, a bulk substrate or a single layer of a material is used as a substrate; for example, glass ($SiO_2$) or $MgF_2$ layer are commonly used as substrates. For an Ag NW on a single layer substrate, there are two supported modes. [13] The first one is a plasmonic bound mode where the field is localized at the nanowire–glass (or $MgF_2$) interface, where the effective refractive index is larger than that of the substrate. The second one is the plasmonic leaky mode where the field extends into both the metal–air and metal–glass interfaces. The propagation constant β for the leaky mode is much smaller than that for the bound mode. The effective index of the NW for the leaky modes is smaller than that of the substrate; therefore, the leaky mode can couple to photons in the substrate. The thinner the Ag NW, the larger the leakage loss, so the plasmonic leaky mode will disappear in the case of very thin Ag NW, such as when the diameter is less than 100 nm (sometimes even below 200 nm) and incident wavelength is in the visible light band.[13-18]   In this paper, we will show that this plasmonic leaky mode can be sustained in the case of thin Ag NW (70 or 90 nm in diameter) if this NW was placed on a dielectric multilayer substrate (DMS) that contains a photonic band gap (PBG), but not the commonly used single layer glass or silicon substrate. And also, the plasmon propagation distance can be increased.

The dielectric multilayer (made of alternating $SiO_2$ and $Si_3N_4$ layers) was fabricated using plasma-enhanced chemical vapour deposition. The thickness of each layer is shown in Figure 1(a). Here, $SiO_2$ is the low (L) refractive index dielectric and $Si_3N_4$ is the high (H) refractive index dielectric. The thicknesses of these dielectric layers were 105 and 88 nm, respectively. In total fourteen layers were deposited on top of each other. The thickness of the top $SiO_2$ layer was approximately 170 nm. By using these structural parameters, this dielectric multilayer can contain the photonic band gap and support two-dimensional Bloch surface waves (2DBSW) at short wavelengths, such as at 590 nm. The 2DBSW can propagate freely across the two dimensional surface, the X-Z plane in Figure 1 (a). When the wavelength is increased (620 nm or longer), the 2DBSW will disappear (See Figure S1). [19] Ag NWs (from Nanjing XFNANO Materials Tech Co., Ltd., China) with a diameter of 90 nm were synthesized using a wet chemistry approach (Figure 1(b)).[1] We deposited an individual NW on the above-mentioned multilayer. A laser beam (from a SuperK supercontinuum source, NKT Photonics, Denmark) was used for excitation; it was first lens-coupled into a standard single mode silica fibre and then directed into the nanofiber through a fibre taper yielding quasi-circular-polarization guided modes with fractional evanescent fields. The silica nanofiber was then brought into contact with the Ag NW (Figure 1(a)) by a micromanipulator. The light propagating along the NW was characterized by a home-built leakage radiation microscope (LRM) system with both a front focal plane (FFP) and back focal plane (BFP) imaging modules.[16]

The FFP image is very common for optical microscope, and provides the spatial distribution of the targets placed on the front focal plane of the objective. Whereas, the BFP imaging module is not always equipped on a commercial microscope. A schematic illustration of the formation of BFP image is shown in Figure 1 (c). At first, we should keep in mind that every spot on the BFP image represents the information of an angle (it can be the leakage radiation angle of the SP, [21]or emitting angle from a dye molecule, [22]). Then, the following will explain how to derive the value of this angle from the corresponding spot on the BFP image. For examples, when two emitters ($S_1$ and $S_2$) with different emission wavelengths were placed on the FFP of an objective, and let's assume their emission is directional and along a particular angle (it is not practical, just an assumption for making an example). These angles into the substrate can be defined as $\vartheta_1$ and $\vartheta_2$. Subsequently, on the BFP of the objective, two spots (P1 and P2) will appear. The radial distance between the two points (P1 and P2) and the centre spot (O) on the BFP is defined as ($r_1$ and $r_2$). Based on the known radius of the *R* (that is determined by the known numerical aperture (NA) of the oil-immersed objective) of the imaging system, and measured $r_1$ and $r_2$ from BFP image, the emitting angles $\vartheta_1$ and $\vartheta_2$ can be derived with the equation *$\vartheta_{1,2}$ = arcsin ((N.A \*$r_{1,2}$)/(R \* n))*, where *n* is the refractive index of the oil or substrate. It is easy to understand that the BFP image can give out information of the angular distribution of the emission or leakage radiation (which is also called as the Fourier plane imaging, meaning the Fourier transformation of the FFP image).

In the following experiment, the LRM is used to measure the SP propagating along an Ag NW placed on the dielectric multilayer. The image formation can be illustrated simply in Figure 1(d). The SP propagates along the NW (*Z*-axis) with wavevector as $k_{sp}$. Due to the leaky property of this plasmonic leaky mode, the SP would become a photon (light) in the multilayer and then coupled into the lower glass substrate (Figure 1 (a)). The magnitude of the wavevector of light (in the glass below the multilayer) parallel to the axis of the NW is $k_z/k_0 = n* \sin(\vartheta) \cos(\phi)$, which should be equal to the $k_{sp}/k_0$, due to the requirement of momentum matching for the excitation and leakage radiation of SP wave, where $k_0 = 2\pi/\lambda$ is the free space wavevector of the light at wavelength $\lambda$. This analysis shows that, on the BFP, the SP would appear as a line perpendicular to the reciprocal space vector corresponding to the long axis of the NW (red dashed line on Figure 1(d)). [13, 17] Using the measured angle $\vartheta$ and $\phi$ by the BFP image (Figure 1 (c) denotes how to derive the angle $\vartheta$ from the BFP image, and $\phi$ is defined on Figure 1 (d) which can be easily derived out), the propagation constant of the plasmonic leaky mode ($k_{sp}/k_0$) can be derived. The schematic of the experimental setup (Figure S2) shows the light paths for both FFP and BFP imaging.

Experimental FFP images of the laser beam propagating along the 90 nm Ag NW are shown in Figure 2(a) and (b) at incident wavelengths of 590 and 630 nm, respectively. Figure 2(c) and (d) shows the laser beam and white light images for an incident wavelength of 660 nm. The length of the Ag NW was approximately 53 μm. The FFP images demonstrate that the plasmonic leaky mode can be sustained on the thin Ag NW on the multilayer substrate at both 590 and 630 nm. The propagation lengths at 590 and 630 nm could be derived with exponential curve fits and were found to be 4 and 16 μm, respectively, based on the intensity profiles along the NW (Figure 2(e) and (f)). The BFP images at these two wavelengths display another difference: at 590 nm a 2DBSW excited by the laser from the fibre taper is visible in the form of bright circles (Figure 2(g)). [21] At a wavelength of 630 nm these circles are not present (Figure 2 (h)). The vertical bright lines on Figure 2(g) and Figure 2(h) are a unique but relatively unrecognized feature of BFP images that correspond to the leaky mode along an Ag NW, [15–18] whose formation is illustrated in Figure 1 (d). These bright lines (labelled "SP") in Figure 2(e) and (f) can be used to derived the effective index of the plasmonic leaky mode (also referred to as the real part of the propagation constant), which was determined to be 1.05 at 590 nm and 1.02 at 630 nm. When the incident wavelength was increased to 660 nm, we did not observe any propagation along the NW in either the FFP (Figure 2(c), and 2(d)) or BFP images (Figure 2(i)). When a NW with the same diameter (90 nm) was placed on a glass substrate, no leaky propagating mode could be observed, similar to that shown in Figure 2(c).

To clearly explain the experimental results, the projected band structure of the dielectric multilayer (shown in Figure 3(a) was calculated for transverse-electric (TE) and transverse-magnetic (TM) waves based on the dispersion equation for Bloch waves) using the finite element method.[22,23] The permittivity of Ag NWs at different wavelengths was set based on experimental values.[24] Owing to the surface scattering and grain boundary effects in real thin films, in the simulation, the refractive indices of $SiO_2$ and $Si_3N_4$ are $n_{SiO2}$ = 1.46 + i × $10^{-3}$ and $n_{Si3N4}$ = 2.14 + i5 × $10^{-3}$, respectively. The refractive index of the glass substrate was $n_{glass}$ = 1.515. It should be noted that generally chemically prepared Ag NWs have pentagonal cross-section, but in our work, this NW is very thin (the diameter is less than 100 nm) and the incident wavelength (around 600 nm) is more than six times of this diameter, so in our simulations, we assume the cross section of the Ag NW is to be circular, which does not affect the modes sustained by the Ag NW.

The dispersion relation for plasmonic leaky mode of Ag NW with a diameter of 90 nm is shown in Figure 3(a) with a red solid line. Based on the electric field distribution and vector directions of electric field on the inset graph of Figure 3 (b), this mode can be attributed to the H1X mode.[25] For an excitation wavelength of 660 nm, the dispersion curve of the H1X mode disappears and the mode is cut off because it moves across the light line (Figure 3 (a)). Therefore, the electromagnetic waves at this frequency cannot be guided by the NW. As noted in Figure 3(a), the dispersion curve of H1X mode is in the stop band for TE polarization. The bandgap effect of dielectric multilayer can inhibit the leakage radiation of H1X mode, and thus increase the propagation distance of H1X mode. Figure 3(b) and 3(c) demonstrate the diameter-dependent dispersion relation of the H1X mode for a NW placed on the dielectric multilayer and on the glass substrate, respectively. It is clearly shown that the H1X mode of Ag NWs placed on the glass substrate has a larger cut-off diameter and shorter propagation distance compared with that of NWs placed on the dielectric multilayer. For examples, when the diameter of the Ag NW is smaller than 480 nm, the H1X mode on the glass substrate is cut off at the incident wavelength of 630 nm. The H1X mode on the dielectric multilayer can also exist even when the diameter of the NW is decreased to 50 nm (as shown by inset graph in Figure 3 (d)). What is more, the comparison between Figure 3 (b) and (c) shows that the propagation distance of the plasmons signals is highly increased if the Ag NW was placed on the DMS.

Here we give an explanation on the comparisons between the plasmonic modes of an Ag NW on a bare glass substrate and that on a DMS. As shown in Figure S3, all the related plasmonic modes of an Ag NW placed on a bare glass substrate were simulated. In Figure S3, there are three plasmonic modes with the incident wavelength fixed at 630 nm. The H0 mode is the plasmonic bound mode with effective refractive index larger than the refractive index of the glass substrate (Figure S3 (a)), although it can sustained at a very thin Ag NW, its propagation distance is much lower (Figure S3 (b)). For example, when the diameter of the Ag NW is 90 nm, the propagation is less than 2 μm (Figure S3 (b)). But this mode has high spatial confinement (Figure S3 (c)). There are another two plasmonic leaky modes, the H1X, and H1Y (Figure S3), they are the typical plasmonic leaky modes which have also been presented in Figure 2d of Reference 18. They are of different polarization states (as noted by the white arrows in Figure S3 (c)). From Figure S3, we also can find these two modes cannot sustain in the case of thin NW and also their propagation distances are much short. For examples, the H1X mode is cutoff when the diameter is less than 440nm, and the H1Y mode is cutoff when the diameter is less than 110 nm. The propagation distance of these two modes are both less than that of the plasmonic mode reported here.

As described in Figure. 1 of Reference 25, there is also a plasmonic mode between the H0 and H1Y, but based on our simulation, we find that this mode will disappear when the incident wavelength is longer than 400 nm (the diameter of the NW is 200 nm, the same as that used in Reference 25). Whereas, in our experiments and simulations, the investigate wavelength is longer than 550 nm, so we do not use this mode for comparison. Based on the polarization direction and electric field distribution, we find that the H1X mode on a glass substrate (Figure S3 (c)) is similar as the plasmonic leaky mode of the Ag NW on the dielectric multilayer (inset graph of Figure 3 (b)). So we also name this plasmonic leaky mode on the DMS that we investigated here as the H1X mode. In the comparison, we have not quantitatively compared the confinement of these plasmonic modes, because it is not precise to define the mode area for a plasmonic leaky mode, and our aim of this paper is mainly focused the longer propagation distance of plasmons sustained on a very thin Ag NW.

Furthly, when we take a close look at the diameter-dependent propagation distances for Ag NWs on a glass substrate (Figure S3 (b)) and that on a DMS (Figure 3 (b)), we can find that the propagation distance decreases with the dimeter of the Ag NWs when the NWs were placed on DMS (Figure 3 (b)), whereas, this dependence is opposite in the case of Ag NWs on a glass (Figure S3).

This phenomenon can be analyzed as following. Considering that the H1X mode is a hybrid mode, [25] the field components of the H1X mode can be decomposed into TE- and TM-polarized components in the dielectric multilayer. The TE- and TM-polarized component of H1X mode can experience difference leaky loss in the dielectric multilayer due to the different photonic band structure for TE and TM polarizations as shown in Figure 3(a). For an excitation wavelength of 630 nm, the dispersion curve of H1X mode is in the stop band for TE polarization. Then, the dielectric multilayer can effectively decrease the leaky radiation of the TE-polarized component of H1X mode based on the photonic bandgap effect. However, the dispersion curve of H1X mode is in the transmission band for TM polarization as noted in Figure 3(a). The TM-polarized component of H1X mode will propagate through the dielectric multilayer and radiate into the glass substrate. To qualitatively understand the wave guiding behaviour of the H1X mode on the dielectric multilayer, the ratio of the electric field energy ($A_x$) with TE-polarized component to the total electric field energy ($A_{all}$) are calculated and is shown in Figure 3(d). $A_x$ and $A_{all}$ are defined as follows: [26]

$$A_x = \int\int_{-\infty}^{+\infty} \frac{1}{2}\left(\frac{d(\varepsilon(r)\omega)}{d\omega}|E_x(r)|^2\right)d^2r$$

$$A_{all} = \int\int_{-\infty}^{+\infty} \frac{1}{2}\left(\frac{d(\varepsilon(r)\omega)}{d\omega}|E(r)|^2\right)d^2r$$

where $|E_x(r)|^2$ and $|E(r)|^2$ are the intensity of the X-component of electric fields and the total electric fields, respectively. $\varepsilon$ is the permittivity. $r$ denotes a spatial position. With the decreasing of the NW diameter, the ratio of $A_x/A_{all}$ is increased, and finally approaches to one. This means that the H1X mode on the dielectric multilayer is gradually transformed into the "pure" TE-polarized mode with decreasing of the NW diameter. For TE-polarized mode, the dielectric multilayer can effectively decrease the leaky radiation of mode due to the bandgap effect. Moreover, as the diameter of NW is decreased, the electric field energy of H1X mode on the dielectric multilayer is increasingly localized into the local environment as noted in the inset of Figure 3(d), which can decrease the loss from the absorption of material and increase the effects of the environment or solution outside the NW. Therefore, the propagation distance of H1X mode on the dielectric multilayer is sharply increased with decreasing of the NW diameter.

In contrast, when the Ag NW was placed on a bare glass substrate, and as the diameter of NW is decreased, the energy of corresponding H1X mode (Figure 3 (c)) is gradually diffused into the environment due to the decreasing of the effective refractive index of mode (Figure 3 (c) and Figure S3). The leakage loss of this H1X mode is enhanced (there is no band gap effect for the bare glass substrate), which decreases the propagation distance of the H1X mode on the bare glass substrate.

In our experiment, an Ag NW of smaller diameter (at 70 nm) placed on the same DMS was also investigated and the results are shown in Figure S 4. The propagation distance can be derived as about 12 μm with the incident wavelength at 590 nm, and 18 μm at 600 nm. These propagation distances are longer than those from the Ag NW with the diameter at 90 nm, which are consistent with the above numerical simulations.

Finally, we will show the polarization state of the H1X mode of an Ag NW on the MDS, and its coupling with another kind of surface wave. When a polarizer was placed before the camera for BFP images, experimental results (Figure S 4 (g) and (f)) show that this H1X mode presents the similar polarization state as the reported BSW mode of a polymeric NW placed on a DMS. [21] This polarization similarity was also verified by the numerical simulation as shown in Figure S5. [21, 27], thus providing the possibility of coupling between the plasmonic mode on an Ag NW and the BSW mode on a polymeric NW. Similar to the Ag NW, the thin polymer NW is also very soft and should also be placed on a solid substrate for practical use. When the polymer NW is too thin and placed on a glass substrate, such as a diameter less than 250 nm, it also cannot transport optical signals. Our previous work and other published paper shows that if we placed this thin polymeric NW on a dielectric multilayer, it can transport optical signals, which is the Bloch surface wave along a dielectric NW or stripe. [21, 27]

Here, to show this coupling, a polymer NW with a diameter of 170 nm was selected. The dispersion relation of the corresponding BSW mode along this polymer NW was simulated as shown in Figure 3 (a) (labelled as BSW-1D). The coupling behaviours between the BSW mode and the plasmonic mode are numerically simulated as shown in Figure S6. The theoretical details of coupling efficiency were shown and discussed in Figure S7. In experiments, a polymer NW with a diameter of about 170 nm was fabricated and placed on the same multilayer substrate (Figure 1(a)); then, an Ag NW (90 nm in diameter) was placed close to the polymer NW (Figure 4(a)). The BSW propagating along the polymer NW was excited via a silica fibre taper, and the BSW was coupled into the SP (Figure 4(b)). The propagation paths of the optical signals along both the polymer and Ag NWs can be clearly observed (Figure 4(b)), which demonstrate that the BSW mode can be coupled into the SP mode. This structure containing two contacted NWs (Figure 4(b)) works as a beam splitter. The original BSW is divided into two paths; the first one sends the BSW along the polymer NW, while the second one changes the BSW into a SP along the Ag NW. Known to all, polymer and metal NWs both have their particular properties, such as polymer NWs can be doped with various functional units, metal Ag NWs might provide high electric field enhancement or sensitivity, so the integration between this two kinds of NWs for signals transportation will open new opportunities for NW devices. Obviously, this kind of modes coupling (between BSWs and SPs on the NWs) is different from the traditional coupling between a dielectric NW and a metal NW, because the NWs on a traditional glass or silicon substrate cannot transport optical signals in the case of a small diameter.

In conclusion, we report the first use of a dielectric multilayer containing a PBG as a substrate for the Ag NW. When comparing with the plasmonic bound mode of an Ag NW on the glass substrate, the plasmonic mode of an Ag NW on the DMS presents much longer propagation distance but less spatial confinement. When comparing with the plasmonic leaky mode of an Ag NW on the glass substrate, the plasmonic mode of the Ag NW on a DMS can be sustained even that the diameter is as low as 70 nm or 90 nm, and keep a longer propagation distance. Although it is a plamonic leaky mode, its propagation distance is still long due to the PBG inside the DMS. This propagation distance can be further increased if the number of the dielectric layer is increased, because that the leaky loss can be further decreased if more dielectric layer was used. But in this case, the propagation of the plasmon cannot be imaged with the LRM. This long propagation distance is favourable for plasmon signals delivery for remote sensing or integrated circuits. As shown in the FFP images, the electric field of this plasmonic mode is located around the side-walls of the NW and propagating much long along the NW, thus providing two parallel localized line-sources with a very small gap (determined by the diameter of the NW), which may be useful for the light-matter interaction in nanometre scale.

This DMS can also tune the polarization state of the plasmonic mode of the Ag NW, so that it can be coupled with the polymer NW in the nanometre scales. Whereas, the optical signals cannot be coupled between these two NWs if they are placed on a commonly glass substrate. Our work will bring new opportunities in the field of nanophotonics because it presents a different platform for the development of NWs devices.[28] The DMS provides new parameters to tune the propagation of an Ag NW Plasmonic waveguide. Therefore, our work will likely lead to the discovery of further intriguing phenomena and the uncovering of further practical applications.

## Supporting Information

The Supporting Information is available free of charge on the ACS Publications website at DOI:XXX.

Figure S1-S7.

**Competing financial interests:** The authors declare no competing financial interests.

## Acknowledgements

This work was supported by MOST (2013CBA01703 and 2016YFA0200601), NSFC (61427818 and 11374286), and the Science and Technological Fund of Anhui Province for Outstanding Youth (1608085J02 and 1608085J01). This work was also supported by grants from the National Institute of Health (GM107986, EB006521, EB018959, and OD019975). This work was partially carried out at the University of Science and Technology of China's Center for Micro and Nanoscale Research and Fabrication. We thank Xiaolei Wen, Linjun Wang, and Yu Wei for their help on the micro/nano fabrication steps. The authors would like to acknowledge Dr Shunping Zhang, Wuhan University, for the very useful discussions on the plasmonic modes of Ag NWs.

## REFERENCES

1. Rycenga, M., Cobley, C.M., Zeng, J., Li, W.Y., Moran, C.H., Zhang, Q., Qin, D., Xia, Y. N., Controlling the Synthesis and Assembly of Silver Nanostructures for Plasmonic Applications. *Chem. Rev.* **2011,** 111, 3669–3712

2. Sun, Y.G., Silver Nanowires Unique Templates for Functional Nanostructures, *Nanoscale,* **2010**, 2,

1626–1642.

3. Guo, X., Ma, Y.G., Wang, Y.P., Tong, L.M., Nanowire Plasmonic Waveguides, Circuits and Devices. *Laser Photonics Rev.* **2013**, 7, 855–881.

4. Sanders, A. W., Routenberg, D. A., Wiley, B.J., Xia, Y.N., Dufresne, E.R., Reed, M.A., Observation of Plasmon Propagation, Redirection, and Fan-out in Silver Nanowires. *Nano. Lett*. **2006,** 6, 1822–1826.

5. Ditlbacher, H., Hohenau, A., Wagner, D., Kreibig, U., Rogers, M., Hofer, F., Aussenegg, F.R., Krenn, J.R., Silver Nanowires as Surface Plasmon Resonators. *Phys. Rev. Lett*. **2005**, 95, 257403.

6. Pyayt, A. L., Wiley, B., Xia, Y. N., Chen, A. T., Dalton, L.，Integration of Photonic and Silver Nanowire Plasmonic Waveguides, *Nat. Nanotechnol.* **2008**, 3, 660-665.

7. Wei, H., Zhang, S.P., Tian, X.R., Xu, H.X., Highly Tunable Propagating Surface Plasmons on Supported Silver Nnanowires. *Proc. Natl. Acad. Sci. USA*, **2013,** 110, 4494–4499.

8. Wei, H., Li, Z.P., Tian, X.R., Wang, Z.X., Cong, F.Z., Liu, N., Zhang, S.P., Nordlander, P., Halas, N.J., Xu. H.X., Quantum Dot-Based Local Field Imaging Reveals Plasmon-Based Interferometric Logic in Silver Nanowire Networks. *Nano Lett*., **2011**, 11, 471–475.

9. Fang, Y.R., Li, Z. P., Huang, Y. Z., Zhang, S.P., Nordlander, P., Halas, N.J., Xu, H.X., Branched Silver Nanowires as Controllable Plasmon Routers. *Nano Lett*. , **2010**, 10, 1950–1954.

10. Wei, H., Wang, Z. X., Tian, X. R., Kall, M., Xu, H.X., Cascaded Logic Gates in Nanophotonic Plasmon Networks. *Nat. Commun.*, **2011**, 2, 387, doi: 10.1038/ncomms1388

11. Wu, X.Q., Xiao, Y., Meng, C., Zhang, X.N., Yu, S.L., Wang, Y.P., Yang, C.X., Guo, X., Ning, C. Z., and Tong, L.M., Hybrid Photon-Plasmon Nanowire Lasers, *Nano Lett*. **2013**, 13, 5654 – 5659.

12. Guo, X., Qiu, M., Bao, J.M., Wiley, B. J., Yang, Q., Zhang, X.N., Ma, Y. G., Yu, H. K., and Tong, L.M., Direct Coupling of Plasmonic and Photonic Nanowires for Hybrid Nanophotonic Components and Circuits, *Nano Lett.*, **2010,** 9, 4515-4519.

13. Johns, P., Beane, G., Yu, K., and Hartland, G. V., Dynamics of Surface Plasmon Polaritons in Metal Nanowires, *J. Phys. Chem. C.*, **2017,** 121, 5445-5459.

14. Zia, R., Selker, M. D., and Brongersma, M. L., Leaky and bound modes of surface plasmon waveguides, *Phys. Rev. B*. **2005**, 71, 165431/1-9

15. Song, M.X, Bouhelier, A., Bramant, P., Sharma, J., Dujardin, E., Zhang, D.G., Colas-des-Francs, G., Imaging Symmetry-Selected Corner Plasmon Modes in Penta-Twinned Crystalline Ag Nano-wires. *ACS Nano*, **2011**, 5, 5874–5880.

16. Yang, H.B., Qiu, M., and Li, Q., Identification and Control of Multiple Leaky Plasmon Modes in Silver Nanowires, *Laser Photonics Rev.* **2016**, 10, 278–286.

17. Wang, Z. X., Wei, H., Pan, D., and Xu, H.X., Controlling the Radiation Direction of Propagating Surface Plasmons on Silver Nanowires, *Laser Photonics Rev*. **2014**, 8, 596–601.

18. Jia, Z. L., Wei, H., Pan, D., and Xu, H.X., Direction-Resolved Radiation from Polarization-Controlled Surface Plasmon Modes on Silver Nanowire Antennas, *Nanoscale*, **2016**, 8, 20118–20124.

19. Zhang, D., Badugu, R., Chen, Y.K., Yu, S.S., Yao,P. J., Wang, P., Ming, H., and Lakowicz, J.R., Back Focal Plane Imaging of Directional Emission from Dye Molecules Coupled to One-dimensional Photonic Crystals. *Nanotechnology*, **2014**, 25, 145202/1-10.

20. Descrovi, E., Barakat, E., Angelini, A., Munzert, P., De Leo, N., Boarino, L., Giorgis, F., and Herzig, H. P., Leakage Radiation Interference Microscopy, *Opt. Lett.,* **2013,** 38, 3374-3376.

21. Wang, R.X., Xia, H. Y., Zhang, D.G., Chen, J. X., Zhu, L. F., Wang, Y., Yang, E.C., Zang, T. Y., Wen, X. L., Zou, G., Wang, P., Ming, H., Badugu, R., Lakowicz, J. R., Bloch Surface Waves Confined in One Dimension with A Single Polymeric Nanofibre, *Nat. Commun*. **2017**, 8:14330 | DOI: 10.1038/ ncomms14330.

22. Yeh, P., Yariv, A., and Hong, C.S., Electromagnetic Propagation in Periodic Stratified Media. I. General Theory, *J. Opt. Soc. Am*. **1977**, 67, 423-438.

23. Joannopoulos, J.D., Johnson, S.G., Winn,.J.N., Meade,.R.D., Photonic Crystals: Molding the Flow of Light, **2008,** Princeton University Press

24. Palik, E. D. and Ghosh, G. Handbook of Optical Constants of Solids, **1998**, Academic Press.

25. Zhang, S.P., and Xu, H.X., Optimizing Substrate-Mediated Plasmon Coupling toward High-Performance Plasmonic Nanowire Waveguides, *ACS Nano*, **2012**, 6, 8128-8135

26. Oulton, R. F., Sorger, V. J., Genov, D. A., Pile, D. F. P. and Zhang, X., A Hybrid Plasmonic Waveguide for Subwavelength Confinement and Long-range Propagation. *Nat. Photonics*, **2008**, 2, 496–500.

27. Descrovi, E., Sfez, T., Quaglio, M., Brunazzo, D., Dominici, L., Michelotti, F., Herzig, H. P., Martin, O. J. F., and Giorgis, F., Guided Bloch Surface Waves on Ultrathin Polymeric Ridges, *Nano Lett.* **2010**, 10, 2087-2091.

28. Yu, L., Barakat, E., Sfez, T., Hvozdara, L., Di Francesco, J., and Herzig, H. P., Manipulating Bloch Surface Waves In 2D: A Platform Concept-Based Fat Lens, *Light: Sci. Appl.* **2014**, 3,124/1-7.

**Figures:**

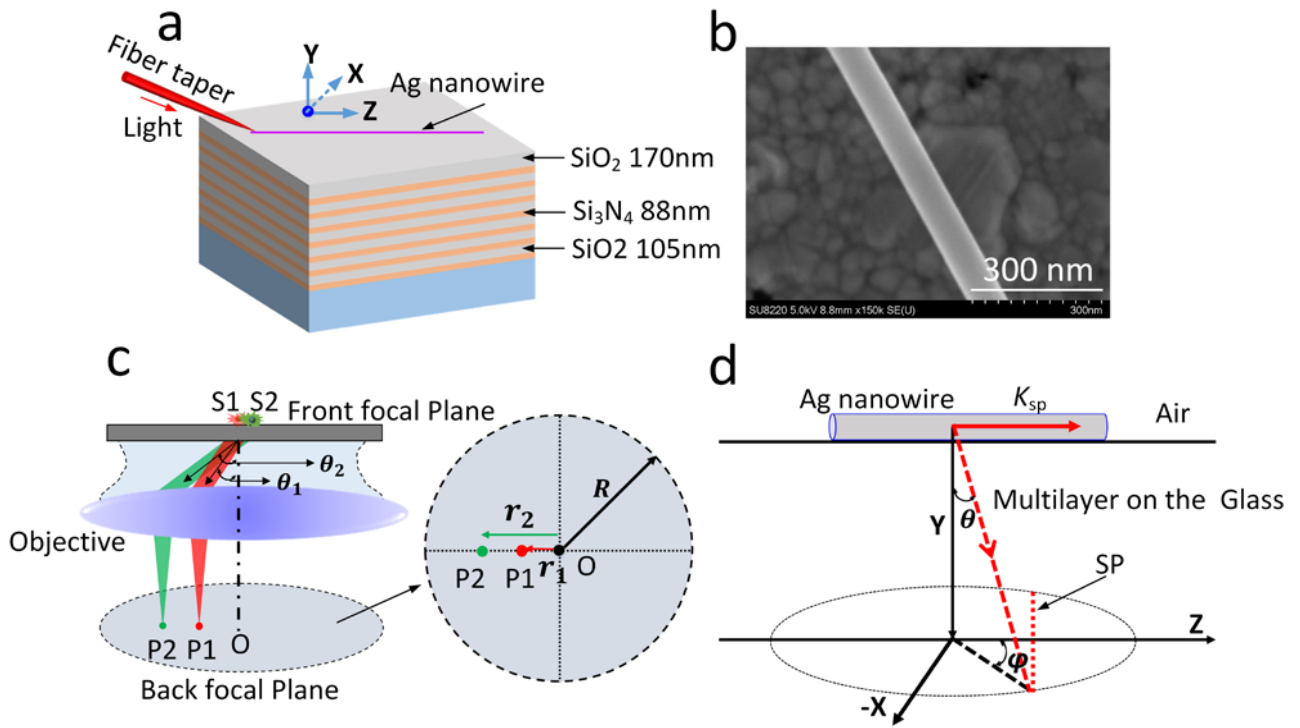

**Figure 1: | Schematic illustration of the experiment.** (a) An Ag NW was placed on a dielectric multilayer consisting of alternating layers of $SiO_2$ (105-nm-thick) and $Si_3N_4$ (88-nm-thick). There were fourteen layers in total, with a top $SiO_2$ layer with a thickness of 170 nm. A fibre taper was used to couple the laser beam into the nanowire. (b) SEM image of an Ag NW (diameter of 90 nm). (c) Illustration of the formation of BFP image. S1 and S2 denote the point emitters which is put on a glass substrate (the front focal plane of the objective). Their emitting angles are defined as $\vartheta_1$ and $\vartheta_2$, then two spots (P1 and P2) appear on the BFP, with distance to the centre spot O as $r_1$ and $r_2$, respectively. (d) Illustration of the BFP imaging formation for the plasmonic leaky mode propagating along an Ag NW. The plasmons with wave-vector ($K_{sp}$) propagate along the NW and leakage radiate simultaneously into the substrate with the angle at ($\vartheta$ and $\phi$). Based on the momentum matching condition, this mode will be represented as a line (red dashed line lab led with SP) on the BFP image.

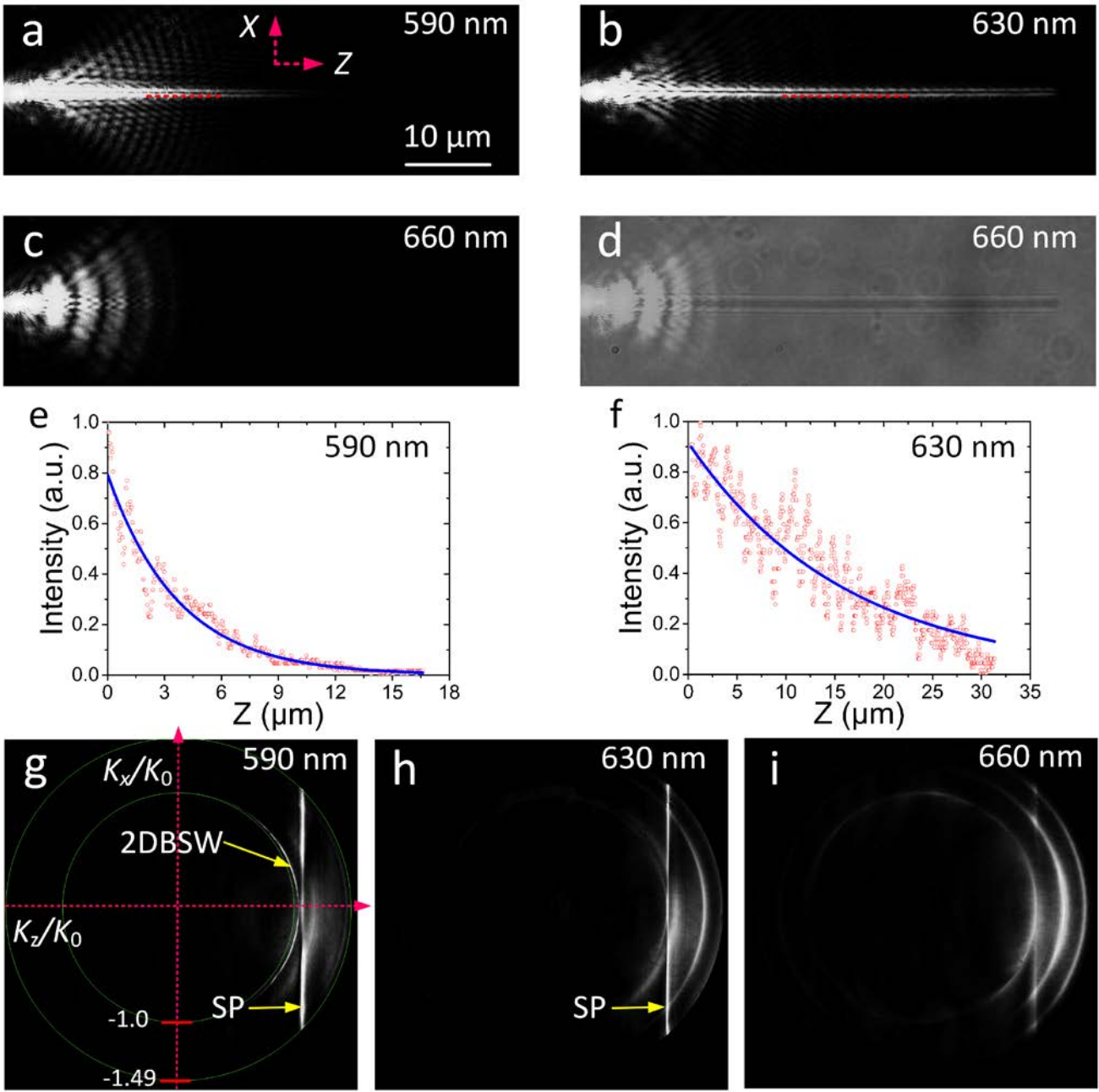

**Figure 2: | Plasmonic mode of the Ag NW on the DMS.** FFP image of the light beam propagating along the Ag NW at three incident wavelengths, namely at (a) 590 nm, (b) 630 nm, and (c) 660 nm. (d) shows the same image area as that in (c), but here a white light source was placed above the sample to illuminate the NW to show the relative position of the coupled fibre taper and the Ag NW which appears as a darker line. The intensity distribution along the NW (Z-direction, red dashed lines) in (a) and (b) are shown in (e) and (f), respectively. The blue solid line is an exponential fit to

the data (red dots) and was used to extract the propagation distance of the plasmonic leaky mode. (g) (h) and (i) show the corresponding BFP images at 590, 630 and 660 nm wavelength, respectively. The scale bar in (a) is also applicable for the images in (b)–(d).

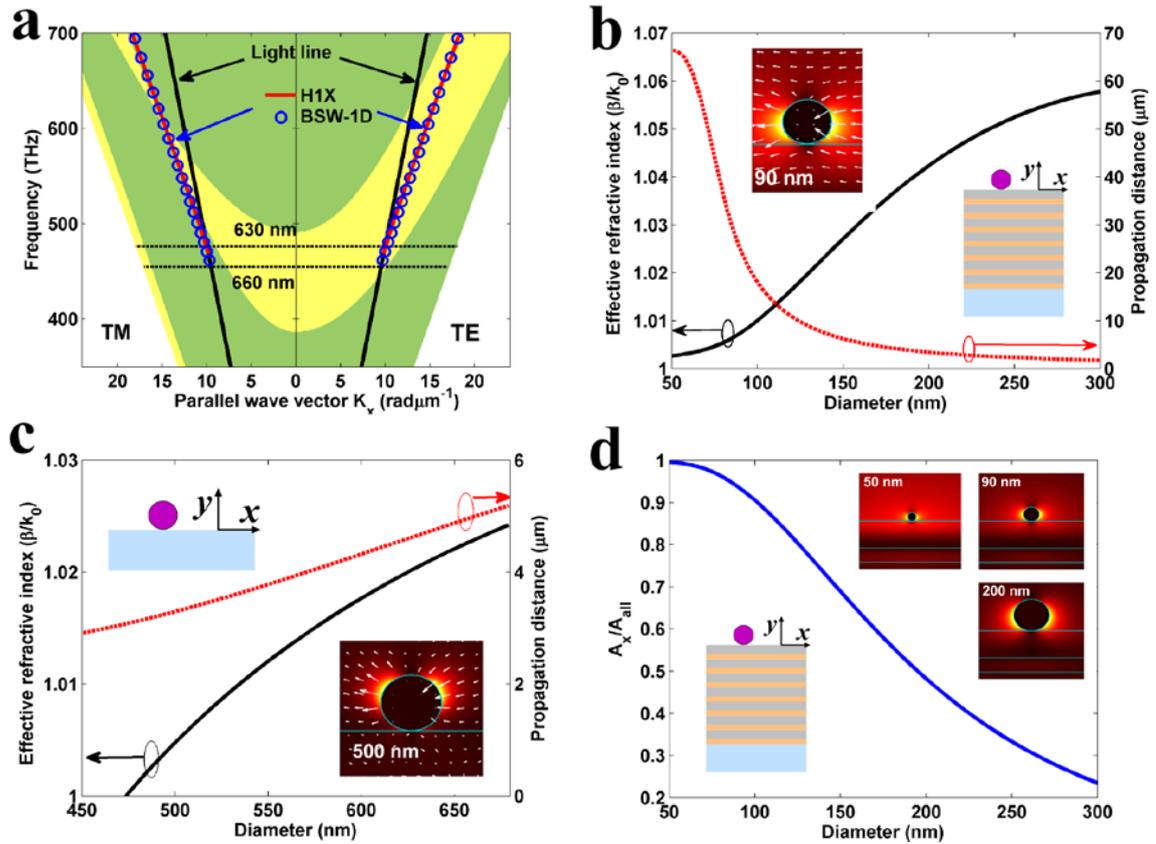

**Figure 3. | Numerical simulation of the plasmonic mode.** (a) The projected band structure of the dielectric multilayer for TE and TM polarizations. The yellow zone denotes the stop band. The red solid lines and blue circles denote the dispersion relation for H1X mode of Ag NW with diameter 90 nm and BSW-1D mode of polymer NW with diameter 170 nm, respectively. (b), (c) show the effective refractive index and propagation distance of the H1X mode versus the diameter of the Ag NW placed on the DMS (b) or on the glass substrate (c). The incident wavelength was 630 nm. The inset in (b) shows the electric field distribution of the H1X mode for the 90-nm-diameter NW on the DMS. The inset in (c) shows the electric field distribution of the H1X mode of the 500-nm-diameter NW on the glass substrate. The arrows in the inset graphs (electric field distribution) denote the direction of the electric field vectors. (d) The ratio of the electric field energy ($A_x/A_{all}$) for the H1X mode versus the diameter. The electric field distributions for H1X mode of Ag NW with diameter 50 nm, 90 nm and 200 nm are shown in the inset, respectively.

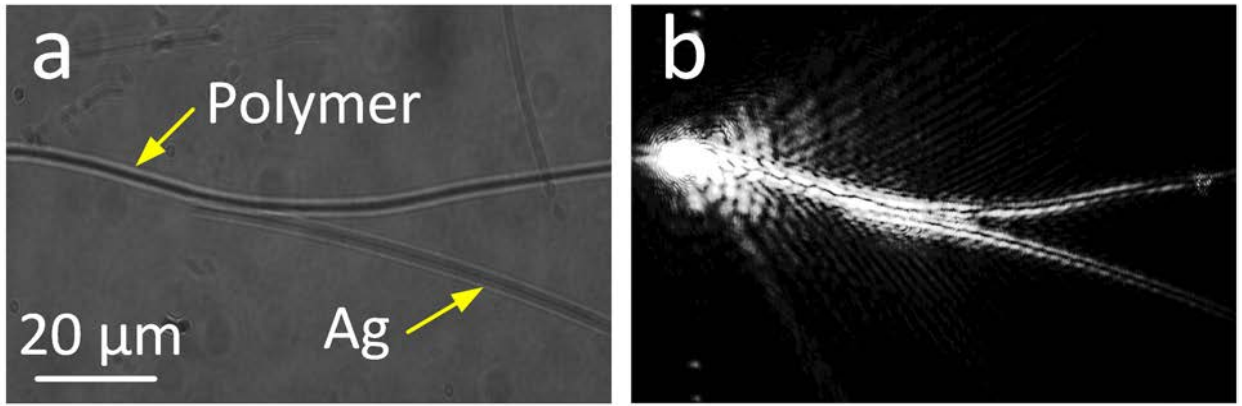

**Figure 4: |Coupling between the plasmonic and BSW modes.** (a) White light image of the Ag NW and polymeric NW on the multilayer. The diameter of the polymer NW is about 170 nm and that of Ag NW is 90 nm. The location of the Ag and polymer nanowires are the darker lines. (b) FFP image of the light beam propagating along the two NWs with the incident wavelength at 630 nm.